\documentclass[a4paper,aps,twocolumn,
balancelastpage,amsmath,,superscriptaddress, amssymb,nofootinbib]{revtex4-1}

\usepackage{setspace}

\usepackage{hyperref}
\usepackage{outlines}
\usepackage{fancyvrb}
\usepackage{amsmath}
\usepackage{hyperref}
\usepackage{braket}
\usepackage{comment}
\usepackage{bbold}
\usepackage{amssymb}
\usepackage{graphicx}
\usepackage[caption=false]{subfig}
\usepackage[usenames, dvipsnames]{color}
\usepackage{soul}

\def\vrms{V_\textrm{RMS}}
\def\vrmsmax{V_\textrm{RMS}^{(\textrm{max})}}

\begin{document}


\title{On intelligent energy harvesting}


\author{Feiyang Liu}
\affiliation{Physics, Southern University of Science and Technology (SUSTech), Shenzhen, China}
\author{Yulong Zhang (co-first)}
\affiliation{EEE, Southern University of Science and Technology (SUSTech), Nanshan District, Shenzhen, China}
\author{Oscar Dahlsten}
\email[]{dahlsten@sustc.edu.cn}
\affiliation{Physics, Southern University of Science and Technology (SUSTech), Shenzhen, China}
\affiliation{Shenzhen Institute for Quantum Science and Engineering, SUSTech, Nanshan District, Shenzhen, China}
\affiliation{London Institute for Mathematical Sciences, Mayfair, 35a South Street, London, W1K 2XF, UK}
\affiliation{Wolfson College, University of Oxford, Oxford OX2 6UD, UK}

\author{Fei Wang (co-corr.)}
\email[]{wangf@sustc.edu.cn}
\affiliation{EEE, Southern University of Science and Technology (SUSTech), Nanshan District, Shenzhen, China}



\date{\today}

\begin{abstract}
 We probe the potential for intelligent intervention to enhance the power output of energy harvesters. We investigate general principles and a case study: a bi-resonant piezo electric harvester. We consider intelligent interventions via pre-programmed reversible energy-conserving operations. We find that in important parameter regimes these can outperform diode-based intervention, which in contrast has a fundamental minimum power dissipation bound. 
\end{abstract}

\pacs{}

\maketitle



{{\bf \em Introduction---}}Energy harvesting, exploiting ambient energy for our purposes, play a crucial role in human technological
development~\cite{Coopersmith10}. Currently, an important focal area is micro energy harvesters (with output power 10-100$\mu$W). These convert, through various transduction methods, ambient thermal and kinetic energy from the environment  to electrical energy. They provide
an in-situ power source for remote electronic devices, typically
for powering sensor nodes of the Internet of Things. This avoids the problems associated with  batteries and/or wiring~\cite{MitchesonYRG08,Kong2014,SujeshaP11,ZhangW18}.

A key challenge for micro harvesters is that ambient energy sources are very often random. For instance, the amplitude and the frequency of a vibrational energy source can be highly variable. This makes it difficult to rectify the generated voltage/current and store the energy in an efficient manner~\cite{MitchesonYRG08,Kong2014, halvorsen08}. 

Interventions by an intelligent agent aids energy harvesting from variable sources in certain contexts, as exemplified by the interventions of a sailor, or a windvane turning a generator into the wind. Such examples serve to remind us that the 2nd law of thermodynamics, as used to prove that a Maxwell's demon cannot work~\cite{Bennett87}, concerns maximum entropy single heat baths, whereas often forces in nature are not maximally random.


Highly sophisticated intelligent interventions in energy harvesting are now practicable, owing to advances in: (i) artificial intelligence software and hardware\cite{BarrancoAS03,LeCunBH15}, (ii) electronic interfacing circuitry~\cite{RamadassC10,SaurayC12,LiangC13,JungmoonK13, Hartmann15}, and (iii) experimental and theoretical understanding of the relation between information and energy, such as the fact that reversible computation has no fundamental energy cost~\cite{Bennett82,sagawa13,JohnH16}. Taken together, this gives significant hope that intelligent intervention may be a powerful tool in mitigating the randomness faced by micro-harvesters.

We therefore here aim to identify intelligent interventions that allow micro harvesters to extract more rectified power from variable sources than current state-of-the art methods. 

\begin{figure}[h]
\includegraphics[width=0.9\linewidth]{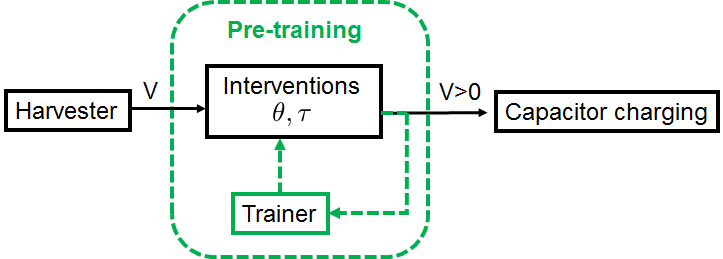}
\caption{We find that rectifying the voltage with intelligently chosen pre-programmed energy conserving interventions can lead to improved power output. Possible interventions include voltage bias flip with ON/OFF-period $\tau$, and voltage phase shift of time $\theta$. The trainer tunes $\theta$ and $\tau$ to optimise the voltage output. }
\label{fig:mainidea}
\end{figure}

A key paradigm we adapt is to use interventions that are reversible and energy conserving. Moreover, for practical and fundamental thermodynamical reasons, these interventions are pre-programmed, chosen by systematic machine learning methods applied to past data from the harvester, as in Fig.\ref{fig:mainidea}. We consider general principles and for concreteness also a case study of a piezo-electric harvester which converts motion to electricity~\cite{LiZ16}. In this case study we consider idealised, pre-programmed, bias-flips and phase-shifts on the electrical outputs. We find these can indeed replace and outperform the current state-of-the art: the diode bridge. 
We moreover note that the the diode bridge has a thermodynamically fundamental lower bound on power dissipation, whereas the methods use here do not. 

 We proceed as follows. We briefly describe the harvester being used as a case study. We describe the interventions, and how they can be intelligently chosen. We then give the results, followed by a discussion and conclusion.

\begin{figure}[h]
\includegraphics[width=0.9\linewidth]{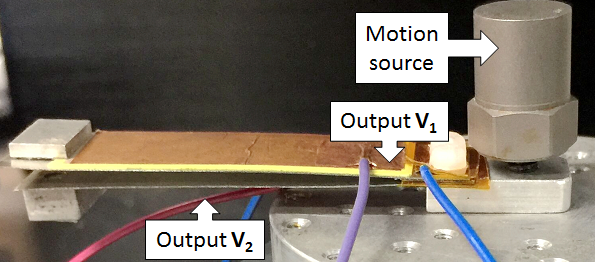}
\caption{The bi-resonant harvester, originally designed in~\cite{LiZ16}. Two piezo-covered cantilevers with masses on their free} ends are driven by the same vibrational motion source on the right axis. We consider how much bias flips and phase shifts on the outputs can enhance the power output.
\label{fig:harvester}
\end{figure}

{{\bf \em Harvester and its output---}}The harvester we use in this paper as a case study is a dual resonant structure energy harvester~\cite{LiZ16}, which can harvest energy from random fluctuation sources at low frequencies (typically less than 100Hz), consistent with motion of everyday objects such as human beings. It consists of two piezoelectric devices, each outputting its own voltage time-series, with the voltages finally combined to give one voltage time-series. The device is shown in FIG.~\ref{fig:harvester}.

The output can be used to charge a capacitor, which is used to power a sensor and  wireless transmission of the sensor signal when required. The capacitor, and normally the sensor and transmission components, needs a DC ($V\geq 0$) source with a sufficiently high root mean square voltage $\vrms$. However the raw voltage from the device is AC. The current state-of-the-art solution for converting it to DC is the {\em diode bridge}.

\begin{figure}[h]
    \centering
    \includegraphics[width=0.6\linewidth]{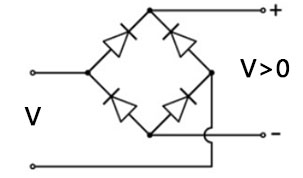}
    \caption{A diode bridge will take any voltage polarity on the inputs to a positive polarity on the output.}
    \label{fig:diodebridge}
\end{figure}

{{\bf \em Diode bridge and its power consumption---}}A diode bridge, as in Fig.\ref{fig:diodebridge}, will take any voltage polarity on the inputs to a positive polarity on the output, but at a loss in power.
The loss in power is necessary given that each diode has a voltage drop. For practical device power dissipation calculations a pn-junction diode's current vs voltage curve can be approximated as $V=V_0+IR$ for the regime $V>0$~\cite{diodepower}. The instantaneous power dissipated by a single diode when $V>0$ is then $P=IV_0+I^2R$. (The average and rms power dissipations follow immediately.) The diode bridge has two diodes in each path and thus twice that dissipation.

Note also that in the case of two sub-harvesters the diode bridge can be applied on each before combining the voltages in order to avoid destructive interference, but again at a power loss.

{{\bf \em 2nd law mandates diode bridge  power consumption---}} If it were possible to reduce the above power dissipation to 0, a diode bridge could be used to violate the second law of thermodynamics, by turning thermal current fluctuations into rectified current at no work cost. Kelvin's version of the 2nd law states that no work can be extracted from a single heat bath in a closed cycle. Thermal voltage fluctuations depend on materials and it is beyond the scope of this paper to investigate their values for diode bridges used here, but e.g.\ the thermal voltage $V_{th}=kT/e$ is approx 0.03V at room temperature. (The argument can be modified to other fluctuation sizes).

For the device to be called a diode the current $ I \approx 0$ for negative voltages beyond the thermal fluctuation range of $-V_{th}$ (up to some breakdown voltage which is outside of the range currently considered). Then for voltages in the positive  thermal fluctuation range we must also have $I\approx 0$, or else the diode would generate a current in a circuit embedded in the heat bath, a circuit which could include a load driven by that current, violating the 2nd law. Thus, according to the above argument, there is an inescapable voltage drop of $V_{th}$ in a diode, and associated power loss. 
We now turn to the competing approach to turn the AC into DC and to remove destructive interference between voltages.

{{\bf \em Two examples of intelligent interventions: sign flip and phase shift---}}The sign flip, which can also be called voltage inversion, can be written as  $V\rightarrow -V$ where $V$ is the instantaneous voltage. This can switch between being on and off with period $\tau$. $\tau$ is a priori a free parameter and will later be set according to optimising based on past data. 

The phase shift is simply a delay of the voltage time series by some amount $\phi$ (so strictly speaking it is a delay rather than a phase shift which should only be between 0 and 2$\pi$ times some period). It can be written as $V(t)\rightarrow V(t+\phi) \, \forall t$ where $t$ is time.

{{\bf \em Interventions are orthogonal matrices---}}
It is convenient to use bra-ket vector notation here such that 
a voltage time series $V_i(t_0), ...V_i(t_f)$ is a vector $\ket{V_i}$
with the first entry $V_i(t_0)$. (The time series is discrete as it is sampled experimentally at a finite rate). The transpose of the vector is denoted $\bra{V_i}$, such that the dot product of two vectors $\ket{V_i}$, $\ket{V_j}$ is written as  $\bra{V_i}\ket{V_j}=\braket{V_i|V_j}.$
In this notation $$ \vrms=\sqrt{\frac{1}{d}\braket{V_i|V_i}}, $$
where $d$ is the dimension of $\ket{V_i}$.
Moreover let $\ket{V'_i}$ denote the transformed $\ket{V_i}$.

In an idealised case the intelligent transformations preserve $\vrms$ such that 
\begin{equation}
\label{eq:Vrmsconstant}
\braket{V_i|V_i}=\braket{V'_i|V'_i}\,\,\forall i. 
\end{equation}
Then, if we also assume the interventions can be represented as matrices, the interventions correspond to orthogonal matrices $O$, meaning $O^TO=I$ where $I$ is the identity and $T$ the transpose. 
The idealised interventions we consider are indeed orthogonal matrices: phase shifting can be represented as a cyclic permutation of elements, a particular permutation matrix, and voltage inversion as a diagonal matrix with diagonal entries all 1 or -1.  
More generally the interventions $\mathcal{S}$ are naturally represented as matrices, since they should respect probabilistic mixtures of different voltages: $\mathcal{S}(\sum_i p_i \ket{V_i})=\sum_i p_i \mathcal{S}(\ket{V_i})$~\cite{Barrett07}. This together with Eq.\ref{eq:Vrmsconstant} implies the idealised interventions, beyond the examples of bias flips and phase shifts, should indeed be represented as orthogonal matrices acting on the voltage vectors. 

{{\bf \em Optimal interventions when combining two voltages---}} Now we can compare the $\vrms$ before and after interventions. For example phase shifts can be used to reduce destructive interference, due to individual sub-generators producing voltages out of phase. Given two or more voltage time series, how high can the $\vrms$ of the combined outputs be, if we are allowed to do intelligent interventions on each time series before combining them? For notational convenience let us consider $d\vrms^2=\braket{V|V}$. Two time series illustrate the general case: $V_1$ and $V_2$. Suppose these undergo the transforms before being combined, how much can 
$d\vrms^2$ change? Note that 
\begin{eqnarray*}
\braket{V'_1\!+\!V'_2|V'_1\!+\!V'_2}-\braket{V_1\!+\!V_2|V_1\!+\!V_2}\!=\\
=\!2(\braket{V'_1|V'_2}\!-\!\braket{V_1|V_2}).
\end{eqnarray*}
Thus maximising the $\vrms$ improvement for a given $\ket{V_1}$, $\ket{V_2}$ means maximising $\braket{V'_1|V'_2}$.  
Can we find a closed form expression for how high this can be? Let us consider maximising it over permutation matrices and sign flips. Note firstly that making the signs the same for all entries, e.g. plus, cannot decrease $\braket{V'_1|V'_2}$. We can assume that in the optimal case the signs are the same, say all positive. Now it is known that the dot product is maximised by ordering the entries of each in descending order: $\braket{V'_1\downarrow|V'_2\downarrow}$.
This follows from the rearrangement inequality. Thus the maximum $d\vrms^2$ one can obtain by signflips and permutations is 
\begin{eqnarray}
\label{Eq:maxvrms}
&\max&_{\mathrm{sgn flip+perms}} \braket{V'_1+V'_2|V'_1+V'_2}\\ \nonumber
&=&\braket{V_1|V_1}+\braket{V_2|V_2}+2\braket{V_1\downarrow|V_2\downarrow}\\ \nonumber
&:=& d{\vrmsmax}^2.
\end{eqnarray}

The operations we consider in this paper, due to engineering considerations, are even more limited: sign-flips and phase shifts (cyclic, not arbitrary permutations).
We therefore do not expect to obtain the $\vrms$ of the closed form expression above but hope to approach it.

{{\bf \em Energy cost of these interventions arbitrarily small---}}
If transformations take individual microstates to other microstates with the same energy they can in principle be performed without an energy cost. Moreover there needs to be a one-to-one mapping between microstates for there not to be a hidden energy cost owing to thermodynamics~\cite{Bennett82}. For example compressing a gas to half its volume isothermally does not change the (average) internal energy of the gas but nevertheless costs work, associated with the reduction of the entropy. 

The interventions here, in idealised form, satisfy those conditions, whereas the diodes do not. The bias flip does not change the potential energy associated with the voltage difference. The phase shift is a delay, again not changing the potential energy. Moreover both operations are reversible, as can be seen physically, and from the fact that orthogonal matrices are reversible.

In contrast, the diode bridge is logically and thermodynamically irreversible with an inescapable lower power dissipation, as discussed above. 

We are investigating what the energetic cost of the interventions will be in practise, taking hope from e.g.\ \cite{RamadassC10} that it can be made low enough to be practical. We also remark here that the interventions are similar to a  feedforward quantum neural net~\cite{WanDKGK17} wherein the transformations are also reversible (unitary), providing another possible physical platform for these ideas.

{{\bf \em Cost function used for the training---}} We wish to optimise the $\vrms$ of the output, under the restriction that it should be DC, i.e. $V>0$. This latter condition is because typically small energy harvesters need to produce DC, e.g. to charge a  capacitor.

For a single voltage the cost function quantifying how far we are from only having positive voltage (POS) can be conveniently implemented as 
\begin{equation}\label{Eq:cpos}
C_{POS}=4\braket{|V|||V|}-\braket{\tilde{V}|\tilde{V}},
\end{equation}
where $\ket{\tilde{V}}=\ket{|V|}+\ket{V}.$ One sees that if all entries are positive, $C=0$, and otherwise $C>0$.

Moreover we define
\begin{eqnarray*}
C_{\vrms} = d{\vrmsmax}^2-d\vrms^2,
\end{eqnarray*}
where $\vrmsmax$ is the maximal over intelligent interventions of equation \ref{Eq:maxvrms}. For simplicity we define the total cost function, taking both desired properties into account, as 
\begin{eqnarray*}
C&=&C_{\vrms}+C_{POS}.
\end{eqnarray*}

An important case here is where the phase shift is done before combining the two voltages, followed by a joint inversion. In this case, in line with Eq.\ref{Eq:maxvrms}, we use the cost function 
\begin{eqnarray*}
C(\tau, \phi)&=&C_{\vrms}+C_{POS}\\
&=&[\bra{V_1\downarrow}\ket{V_2\downarrow}-\\
&\bra{V'_1}&\ket{V'_2}]+ [4\bra{|V'|}\ket{|V'|}-\bra{\tilde{V'}}\ket{\tilde{V'}}],
\end{eqnarray*}
where $\ket{V'}$ is the sum of the two voltages after the first phase shift and $\ket{\tilde{V'}}=\ket{|V'|}+\ket{V'}.$

{{\bf \em Systematic training methods exist---}} To have a systematic and scalable approach we consider the well-proven machine learning/optimisation technique of gradient descent on a suitably defined cost function.
The gradient descent rule as applied here is that
  \[\begin{pmatrix}\tau \\ \theta \end{pmatrix}\rightarrow \begin{pmatrix}\tau \\ \theta \end{pmatrix}-\eta\begin{pmatrix}\frac{C(\theta,\tau+\delta)-C(\theta,\tau)}{\delta} \\ \frac{C(\theta+\delta,\tau)-C(\theta,\tau)}{\delta} \end{pmatrix}, \]
where $\delta$ and $\eta$ are numerical parameters chosen according to what works. 

Moreover, when faced with local minima in the cost function landscape, we employ the genetic algorithm, a type of evolutionary algorithm commonly used to find global minima when there are many local minima. In the genetic algorithm, the global minimum (highest fitness generation) can often be found, after operations like mutation, crossover and selection~\cite{Holland75}.

We use some of the time-series data (80$\%$) for determining the optimal interventions, and then test those interventions on the remaining data (20$\%$). 

The training used here can be classified as {\em reinforcement learning}, as the performance is evaluated (rather than the output being compared to a a known correct answer as in supervised learning).

{{\bf \em Simulated intelligent intervention beats diode bridge---}} Our simulation shows that a combination of periodic voltage inversion and phase shift provides DC voltage that is higher than that after the diode bridge.




The diode bridge penalty of about 0.2V is significant in regimes where $\vrms$ is of the order of 0.2 or less. It is in these regimes it makes sense to consider replacing the diode bridge. 

\begin{figure}
    \centering
    \includegraphics[width=0.8\linewidth]{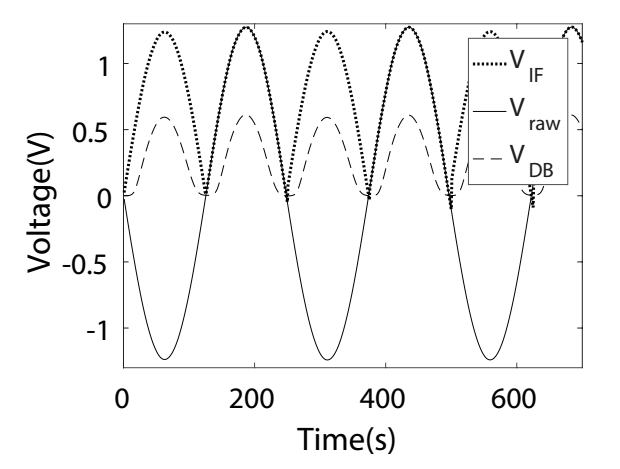}
    \caption{Intelligently chosen periodic voltage inversion gives better $\vrms$ performance than diode bridge for the same input. The diode bridge data is fully experimental. The intelligent intervention data is from applying the corresponding orthogonal matrix on the raw data experimental data before the diode bridge.
    $V_{IF}$ is the voltage after the intelligently chosen flip (experiment+simulated intervention), $V_{\mathrm{RAW}}$ (experimental) is the direct output from the harvester, and $V_{\mathrm{DB}}$ is that after the diode bridge (experimental).  }
    \label{fig:IFbeatsDB}
\end{figure}

Fig.\ref{fig:IFbeatsDB} shows how the $\vrms$ is left essentially undiminished and approximately non-negative by an intelligently chosen periodic voltage inversion, whereas the diode bridge loses about half of the $\vrms$.  
In regimes of even lower $\vrms$ this advantage will be even greater of course. 

\begin{figure}
\includegraphics[width=0.8\linewidth]{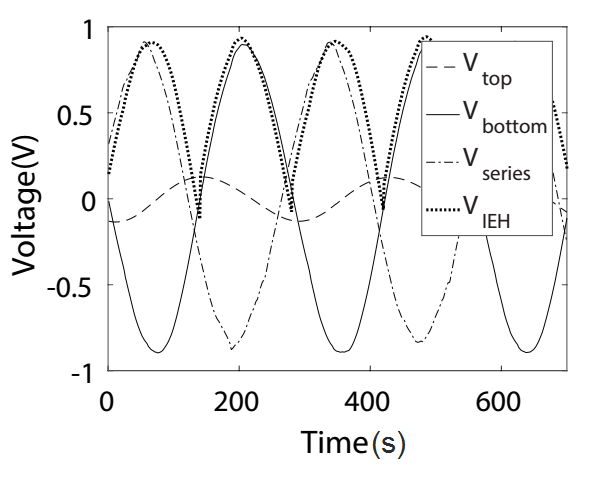}
\caption{Two devices driven by the same source can have very similar frequency but be out of phase, as in this experimental data based on 2 piezo-electric harvesters. The phase difference leads to negative interference. The output of intelligent energy harvesting reduces the consumption of voltage and maximizes the RMS. }
\label{fig:2Voltages}
\end{figure}

\begin{figure}
    \centering
    \includegraphics[width=\linewidth]{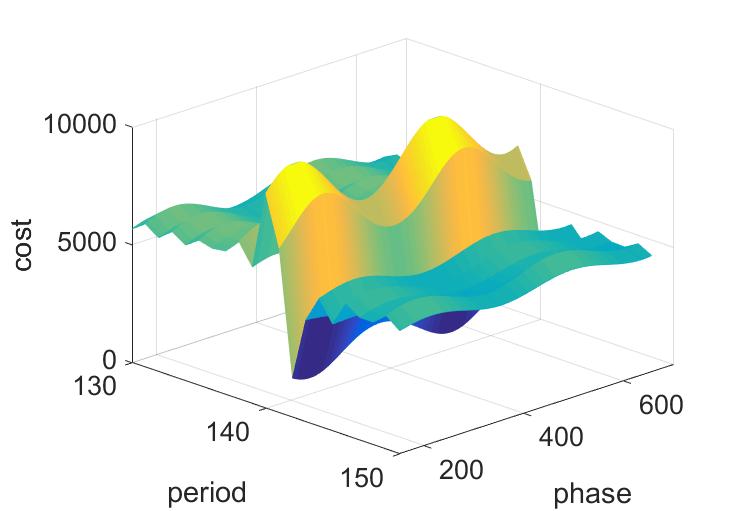}
    \caption{Cost function landscape. X-axis  is period and Y-axis is delay(length of phase shift), Z-axis is the cost function value, which combines a cost for being less than 0 and a cost for having suboptimal Vrms.}
    \label{fig:cost1}
\end{figure}

In the case of two sub-generators we find that the simulated intelligently chosen delay plus intelligently chosen periodic inversion can also in principle significantly outperform the diode bridge, as shown in Fig.~\ref{fig:2Voltages}.
An example of the cost function landscape for real experimental data combined with simulated intervention is in Fig.~\ref{fig:cost1}, showing local minima, which is why we employed the genetic algorithm for the training. 
The $\vrms$ improvement here is given in Table~\ref{Table1}. 
\begin{table}
\renewcommand{\arraystretch}{1.3}
  \centering
    \caption{Improvement of $\vrms$ ($C_{POS}$ of Eq.\ref{Eq:cpos}). Three cases:(i) raw data without intelligent intervention or diode bridge, (ii) data after diode bridge, (iii) data for intelligent intervention replacing diode bridge. For the case of 1 voltage only being used the bias flip is employed. For the case of both voltages being used, there is a phase shift applied, followed by combining the voltages, followed by a bias flip. }
    \label{Table1}
\begin{tabular}{l|c|c}
   $\vrms(\frac{C_{POS}}{L})$  &1 voltage & 2 voltages  \\
\hline
  raw data & 0.89(1.6) & 0.59(0.7)\\ 
  DB & 0.37(0.5) & 0.22(0.2)    \\
  IEH & 0.89(0.01) & 0.59(0.47)   \\
\end{tabular}
\end{table}

{{\bf \em Signal-to-noise ratio important---}} We consider adjusting the power of additive Gaussian white noise to see the change of $\vrms$ and $C$ of the intelligent interventions and and diode bridge respectively.

\begin{figure}
\includegraphics[width=0.9\linewidth]{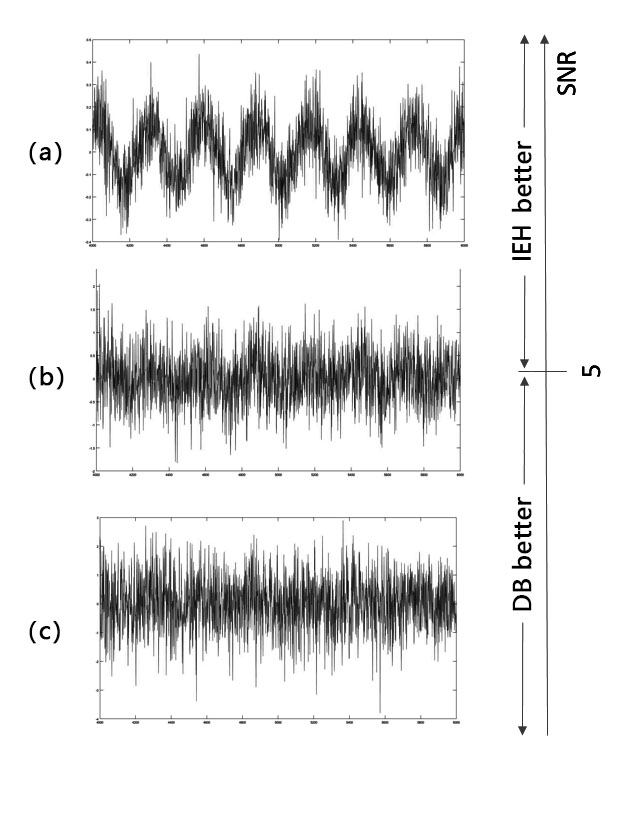}
\caption{Comparison of diode bridge and intelligent intervention under noise with three cases illustrating the corresponding parameter regime: (a) SNR(signal-to-noise ratio) is greater than 5, in which the intelligent harvesting (IEH) has a better cost function performance, (b) SNR  is close to 5, wherein the cost of IEH and the diode bridge (DB) are similar, and (c) SNR  is less than 5, wherein the cost of the diode bridge is lower. }
\label{fig:SNRcom}
\end{figure}

We find that whilst the $\vrms$ of the intelligent interventions is always greater than the diode bridge, when it comes to the total cost function $C$, it looks like in Fig.  \ref{fig:SNRcom}: there is a threshold signal-to-noise ratio SNR after which the diode bridge wins.
This is consistent with the understanding that the intelligent intervention relies on patterns existing.

\vspace{8cm}

{{\bf \em Conclusion---}} We conclude from this study that in the case of small voltages and multiple sub-generators, intelligent intervention can significantly outperform the diode bridge. This suggest it may be strikingly useful to divide existing harvesters into independently moving sub-components together with intelligent interventions on the outputs. Such a harvester with multiple (in the case here 2) sub-harvesters could have a much wider operating range than a single large harvester which may not move at all when exposed to small and/or in different locations opposing forces.

\bibliography{IEHTriboRefs}

\begin{thebibliography}{22}%
\makeatletter
\providecommand \@ifxundefined [1]{%
 \@ifx{#1\undefined}
}%
\providecommand \@ifnum [1]{%
 \ifnum #1\expandafter \@firstoftwo
 \else \expandafter \@secondoftwo
 \fi
}%
\providecommand \@ifx [1]{%
 \ifx #1\expandafter \@firstoftwo
 \else \expandafter \@secondoftwo
 \fi
}%
\providecommand \natexlab [1]{#1}%
\providecommand \enquote  [1]{``#1''}%
\providecommand \bibnamefont  [1]{#1}%
\providecommand \bibfnamefont [1]{#1}%
\providecommand \citenamefont [1]{#1}%
\providecommand \href@noop [0]{\@secondoftwo}%
\providecommand \href [0]{\begingroup \@sanitize@url \@href}%
\providecommand \@href[1]{\@@startlink{#1}\@@href}%
\providecommand \@@href[1]{\endgroup#1\@@endlink}%
\providecommand \@sanitize@url [0]{\catcode `\\12\catcode `\$12\catcode
  `\&12\catcode `\#12\catcode `\^12\catcode `\_12\catcode `\%12\relax}%
\providecommand \@@startlink[1]{}%
\providecommand \@@endlink[0]{}%
\providecommand \url  [0]{\begingroup\@sanitize@url \@url }%
\providecommand \@url [1]{\endgroup\@href {#1}{\urlprefix }}%
\providecommand \urlprefix  [0]{URL }%
\providecommand \Eprint [0]{\href }%
\providecommand \doibase [0]{http://dx.doi.org/}%
\providecommand \selectlanguage [0]{\@gobble}%
\providecommand \bibinfo  [0]{\@secondoftwo}%
\providecommand \bibfield  [0]{\@secondoftwo}%
\providecommand \translation [1]{[#1]}%
\providecommand \BibitemOpen [0]{}%
\providecommand \bibitemStop [0]{}%
\providecommand \bibitemNoStop [0]{.\EOS\space}%
\providecommand \EOS [0]{\spacefactor3000\relax}%
\providecommand \BibitemShut  [1]{\csname bibitem#1\endcsname}%
\let\auto@bib@innerbib\@empty
\bibitem [{\citenamefont {Coopersmith}(2010)}]{Coopersmith10}%
  \BibitemOpen
  \bibfield  {author} {\bibinfo {author} {\bibfnamefont {J.}~\bibnamefont
  {Coopersmith}},\ }\href@noop {} {\emph {\bibinfo {title} {Energy, the subtle
  concept : the discovery of Feynman's blocks from Leibniz to Einstein}}}\
  (\bibinfo  {publisher} {Oxford University Press Oxford},\ \bibinfo {year}
  {2010})\ pp.\ \bibinfo {pages} {xiv, 400}\BibitemShut {NoStop}%
\bibitem [{\citenamefont {Mitcheson}\ \emph {et~al.}(2008)\citenamefont
  {Mitcheson}, \citenamefont {Yeatman}, \citenamefont {Rao}, \citenamefont
  {Holmes},\ and\ \citenamefont {Green}}]{MitchesonYRG08}%
  \BibitemOpen
  \bibfield  {author} {\bibinfo {author} {\bibfnamefont {P.~D.}\ \bibnamefont
  {Mitcheson}}, \bibinfo {author} {\bibfnamefont {E.~M.}\ \bibnamefont
  {Yeatman}}, \bibinfo {author} {\bibfnamefont {G.~K.}\ \bibnamefont {Rao}},
  \bibinfo {author} {\bibfnamefont {A.~S.}\ \bibnamefont {Holmes}}, \ and\
  \bibinfo {author} {\bibfnamefont {T.~C.}\ \bibnamefont {Green}},\ }\href
  {\doibase 10.1109/JPROC.2008.927494} {\bibfield  {journal} {\bibinfo
  {journal} {Proceedings of the IEEE}\ }\textbf {\bibinfo {volume} {96}},\
  \bibinfo {pages} {1457} (\bibinfo {year} {2008})}\BibitemShut {NoStop}%
\bibitem [{\citenamefont {et~al}(2014)}]{Kong2014}%
  \BibitemOpen
  \bibfield  {author} {\bibinfo {author} {\bibfnamefont {K.~L.~B.}\
  \bibnamefont {et~al}},\ }\href@noop {} {\emph {\bibinfo {title} {Waste Energy
  Harvesting: Mechanical and Thermal Energies}}}\ (\bibinfo  {publisher}
  {Springer},\ \bibinfo {year} {2014})\BibitemShut {NoStop}%
\bibitem [{\citenamefont {Sudevalayam}\ and\ \citenamefont
  {Kulkarni}(2011)}]{SujeshaP11}%
  \BibitemOpen
  \bibfield  {author} {\bibinfo {author} {\bibfnamefont {S.}~\bibnamefont
  {Sudevalayam}}\ and\ \bibinfo {author} {\bibfnamefont {P.}~\bibnamefont
  {Kulkarni}},\ }\href@noop {} {\bibfield  {journal} {\bibinfo  {journal} {IEEE
  Communications Surveys \& Tutorials}\ }\textbf {\bibinfo {volume} {13}},\
  \bibinfo {pages} {443} (\bibinfo {year} {2011})}\BibitemShut {NoStop}%
\bibitem [{\citenamefont {Zhang}\ \emph {et~al.}(2018)\citenamefont {Zhang},
  \citenamefont {Wang}, \citenamefont {Luo}, \citenamefont {Hu}, \citenamefont
  {Li},\ and\ \citenamefont {Wang}}]{ZhangW18}%
  \BibitemOpen
  \bibfield  {author} {\bibinfo {author} {\bibfnamefont {Y.}~\bibnamefont
  {Zhang}}, \bibinfo {author} {\bibfnamefont {T.}~\bibnamefont {Wang}},
  \bibinfo {author} {\bibfnamefont {A.}~\bibnamefont {Luo}}, \bibinfo {author}
  {\bibfnamefont {Y.}~\bibnamefont {Hu}}, \bibinfo {author} {\bibfnamefont
  {X.}~\bibnamefont {Li}}, \ and\ \bibinfo {author} {\bibfnamefont
  {F.}~\bibnamefont {Wang}},\ }\href@noop {} {\bibfield  {journal} {\bibinfo
  {journal} {Applied Energy}\ }\textbf {\bibinfo {volume} {212}},\ \bibinfo
  {pages} {362} (\bibinfo {year} {2018})}\BibitemShut {NoStop}%
\bibitem [{\citenamefont {Halvorsen}(2008)}]{halvorsen08}%
  \BibitemOpen
  \bibfield  {author} {\bibinfo {author} {\bibfnamefont {E.}~\bibnamefont
  {Halvorsen}},\ }\href@noop {} {\bibfield  {journal} {\bibinfo  {journal}
  {Journal of Microelectromechanical Systems}\ }\textbf {\bibinfo {volume}
  {17}},\ \bibinfo {pages} {1061} (\bibinfo {year} {2008})}\BibitemShut
  {NoStop}%
\bibitem [{\citenamefont {Bennett}(1987)}]{Bennett87}%
  \BibitemOpen
  \bibfield  {author} {\bibinfo {author} {\bibfnamefont {C.~H.}\ \bibnamefont
  {Bennett}},\ }\href@noop {} {\bibfield  {journal} {\bibinfo  {journal}
  {Scientific American}\ }\textbf {\bibinfo {volume} {257}},\ \bibinfo {pages}
  {108} (\bibinfo {year} {1987})}\BibitemShut {NoStop}%
\bibitem [{\citenamefont {Linares-Barranco}\ \emph {et~al.}(2003)\citenamefont
  {Linares-Barranco}, \citenamefont {Andreou}, \citenamefont {Indiveri},\ and\
  \citenamefont {Shibata}}]{BarrancoAS03}%
  \BibitemOpen
  \bibfield  {author} {\bibinfo {author} {\bibfnamefont {B.}~\bibnamefont
  {Linares-Barranco}}, \bibinfo {author} {\bibfnamefont {A.~G.}\ \bibnamefont
  {Andreou}}, \bibinfo {author} {\bibfnamefont {G.}~\bibnamefont {Indiveri}}, \
  and\ \bibinfo {author} {\bibfnamefont {T.}~\bibnamefont {Shibata}},\ }\href
  {\doibase 10.1109/TNN.2003.819420} {\bibfield  {journal} {\bibinfo  {journal}
  {IEEE Transactions on Neural Networks}\ }\textbf {\bibinfo {volume} {14}},\
  \bibinfo {pages} {976} (\bibinfo {year} {2003})}\BibitemShut {NoStop}%
\bibitem [{\citenamefont {LeCun}\ \emph {et~al.}(2015)\citenamefont {LeCun},
  \citenamefont {Bengio},\ and\ \citenamefont {Hinton}}]{LeCunBH15}%
  \BibitemOpen
  \bibfield  {author} {\bibinfo {author} {\bibfnamefont {Y.}~\bibnamefont
  {LeCun}}, \bibinfo {author} {\bibfnamefont {Y.}~\bibnamefont {Bengio}}, \
  and\ \bibinfo {author} {\bibfnamefont {G.}~\bibnamefont {Hinton}},\ }\href
  {http://dx.doi.org/10.1038/nature14539} {\bibfield  {journal} {\bibinfo
  {journal} {Nature}\ }\textbf {\bibinfo {volume} {521}} (\bibinfo {year}
  {2015})}\BibitemShut {NoStop}%
\bibitem [{\citenamefont {Ramadass}\ and\ \citenamefont
  {Chandrakasan}(2010)}]{RamadassC10}%
  \BibitemOpen
  \bibfield  {author} {\bibinfo {author} {\bibfnamefont {Y.}~\bibnamefont
  {Ramadass}}\ and\ \bibinfo {author} {\bibfnamefont {A.}~\bibnamefont
  {Chandrakasan}},\ }\href@noop {} {\bibfield  {journal} {\bibinfo  {journal}
  {Solid-State Circuits, IEEE Journal Of}\ }\textbf {\bibinfo {volume} {45}},\
  \bibinfo {pages} {189} (\bibinfo {year} {2010})}\BibitemShut {NoStop}%
\bibitem [{\citenamefont {Bandyopadhyay}\ and\ \citenamefont
  {Chandrakasan}(2012)}]{SaurayC12}%
  \BibitemOpen
  \bibfield  {author} {\bibinfo {author} {\bibfnamefont {S.}~\bibnamefont
  {Bandyopadhyay}}\ and\ \bibinfo {author} {\bibfnamefont {A.~P.}\ \bibnamefont
  {Chandrakasan}},\ }\href@noop {} {\bibfield  {journal} {\bibinfo  {journal}
  {IEEE Journal of Solid-State Circuits}\ }\textbf {\bibinfo {volume} {47}},\
  \bibinfo {pages} {2199} (\bibinfo {year} {2012})}\BibitemShut {NoStop}%
\bibitem [{\citenamefont {Liang}\ and\ \citenamefont {Chung}(2013)}]{LiangC13}%
  \BibitemOpen
  \bibfield  {author} {\bibinfo {author} {\bibfnamefont {J.}~\bibnamefont
  {Liang}}\ and\ \bibinfo {author} {\bibfnamefont {H.~S.-H.}\ \bibnamefont
  {Chung}},\ }\href {http://stacks.iop.org/1742-6596/476/i=1/a=012025}
  {\bibfield  {journal} {\bibinfo  {journal} {Journal of Physics: Conference
  Series}\ }\textbf {\bibinfo {volume} {476}},\ \bibinfo {pages} {012025}
  (\bibinfo {year} {2013})}\BibitemShut {NoStop}%
\bibitem [{\citenamefont {Kim}\ and\ \citenamefont {Kim}(2013)}]{JungmoonK13}%
  \BibitemOpen
  \bibfield  {author} {\bibinfo {author} {\bibfnamefont {J.}~\bibnamefont
  {Kim}}\ and\ \bibinfo {author} {\bibfnamefont {C.}~\bibnamefont {Kim}},\
  }\href@noop {} {\bibfield  {journal} {\bibinfo  {journal} {IEEE Transactions
  on Power Electronics}\ }\textbf {\bibinfo {volume} {28}},\ \bibinfo {pages}
  {3827} (\bibinfo {year} {2013})}\BibitemShut {NoStop}%
\bibitem [{\citenamefont {Hartmann}\ \emph {et~al.}(2015)\citenamefont
  {Hartmann}, \citenamefont {Pfeffer}, \citenamefont {H\"ofling}, \citenamefont
  {Kamp},\ and\ \citenamefont {Worschech}}]{Hartmann15}%
  \BibitemOpen
  \bibfield  {author} {\bibinfo {author} {\bibfnamefont {F.}~\bibnamefont
  {Hartmann}}, \bibinfo {author} {\bibfnamefont {P.}~\bibnamefont {Pfeffer}},
  \bibinfo {author} {\bibfnamefont {S.}~\bibnamefont {H\"ofling}}, \bibinfo
  {author} {\bibfnamefont {M.}~\bibnamefont {Kamp}}, \ and\ \bibinfo {author}
  {\bibfnamefont {L.}~\bibnamefont {Worschech}},\ }\href {\doibase
  10.1103/PhysRevLett.114.146805} {\bibfield  {journal} {\bibinfo  {journal}
  {Phys. Rev. Lett.}\ }\textbf {\bibinfo {volume} {114}},\ \bibinfo {pages}
  {146805} (\bibinfo {year} {2015})}\BibitemShut {NoStop}%
\bibitem [{\citenamefont {Bennett}(1982)}]{Bennett82}%
  \BibitemOpen
  \bibfield  {author} {\bibinfo {author} {\bibfnamefont {C.~H.}\ \bibnamefont
  {Bennett}},\ }\href@noop {} {\ \textbf {\bibinfo {volume} {21}},\ \bibinfo
  {pages} {905} (\bibinfo {year} {1982})}\BibitemShut {NoStop}%
\bibitem [{\citenamefont {Sagawa}(2013)}]{sagawa13}%
  \BibitemOpen
  \bibfield  {author} {\bibinfo {author} {\bibfnamefont {T.}~\bibnamefont
  {Sagawa}},\ }in\ \href@noop {} {\emph {\bibinfo {booktitle} {Lectures on
  Quantum Computing, Thermodynamics and Statistical Physics}}}\ (\bibinfo
  {publisher} {World Scientific},\ \bibinfo {year} {2013})\ pp.\ \bibinfo
  {pages} {125--190}\BibitemShut {NoStop}%
\bibitem [{\citenamefont {Goold}\ \emph {et~al.}(2016)\citenamefont {Goold},
  \citenamefont {Huber}, \citenamefont {Riera}, \citenamefont {del Rio},\ and\
  \citenamefont {Skrzypczyk}}]{JohnH16}%
  \BibitemOpen
  \bibfield  {author} {\bibinfo {author} {\bibfnamefont {J.}~\bibnamefont
  {Goold}}, \bibinfo {author} {\bibfnamefont {M.}~\bibnamefont {Huber}},
  \bibinfo {author} {\bibfnamefont {A.}~\bibnamefont {Riera}}, \bibinfo
  {author} {\bibfnamefont {L.}~\bibnamefont {del Rio}}, \ and\ \bibinfo
  {author} {\bibfnamefont {P.}~\bibnamefont {Skrzypczyk}},\ }\href@noop {}
  {\bibfield  {journal} {\bibinfo  {journal} {Journal of Physics A:
  Mathematical and Theoretical}\ }\textbf {\bibinfo {volume} {49}},\ \bibinfo
  {pages} {143001} (\bibinfo {year} {2016})}\BibitemShut {NoStop}%
\bibitem [{\citenamefont {Li}\ \emph {et~al.}(2016)\citenamefont {Li},
  \citenamefont {Peng}, \citenamefont {Zhang},\ and\ \citenamefont
  {Wang}}]{LiZ16}%
  \BibitemOpen
  \bibfield  {author} {\bibinfo {author} {\bibfnamefont {S.}~\bibnamefont
  {Li}}, \bibinfo {author} {\bibfnamefont {Z.}~\bibnamefont {Peng}}, \bibinfo
  {author} {\bibfnamefont {A.}~\bibnamefont {Zhang}}, \ and\ \bibinfo {author}
  {\bibfnamefont {F.}~\bibnamefont {Wang}},\ }\href {\doibase
  10.1063/1.4941353} {\bibfield  {journal} {\bibinfo  {journal} {AIP Advances}\
  }\textbf {\bibinfo {volume} {6}},\ \bibinfo {pages} {015019} (\bibinfo {year}
  {2016})}\BibitemShut {NoStop}%
\bibitem [{\citenamefont {note}(2011)}]{diodepower}%
  \BibitemOpen
  \bibfield  {author} {\bibinfo {author} {\bibfnamefont {S.~S.~A.}\
  \bibnamefont {note}},\ }\href@noop {} {\enquote {\bibinfo {title}
  {Calculation of power loss in a diode},}\ } (\bibinfo {year}
  {2011})\BibitemShut {NoStop}%
\bibitem [{\citenamefont {Barrett}(2007)}]{Barrett07}%
  \BibitemOpen
  \bibfield  {author} {\bibinfo {author} {\bibfnamefont {J.}~\bibnamefont
  {Barrett}},\ }\href {\doibase 10.1103/PhysRevA.75.032304} {\bibfield
  {journal} {\bibinfo  {journal} {Phys. Rev. A}\ }\textbf {\bibinfo {volume}
  {75}},\ \bibinfo {pages} {032304} (\bibinfo {year} {2007})}\BibitemShut
  {NoStop}%
\bibitem [{\citenamefont {{Wan}}\ \emph {et~al.}(2017)\citenamefont {{Wan}},
  \citenamefont {{Dahlsten}}, \citenamefont {{Kristj{\'a}nsson}}, \citenamefont
  {{Gardner}},\ and\ \citenamefont {{Kim}}}]{WanDKGK17}%
  \BibitemOpen
  \bibfield  {author} {\bibinfo {author} {\bibfnamefont {K.~H.}\ \bibnamefont
  {{Wan}}}, \bibinfo {author} {\bibfnamefont {O.}~\bibnamefont {{Dahlsten}}},
  \bibinfo {author} {\bibfnamefont {H.}~\bibnamefont {{Kristj{\'a}nsson}}},
  \bibinfo {author} {\bibfnamefont {R.}~\bibnamefont {{Gardner}}}, \ and\
  \bibinfo {author} {\bibfnamefont {M.~S.}\ \bibnamefont {{Kim}}},\ }\href
  {\doibase 10.1038/s41534-017-0032-4} {\bibfield  {journal} {\bibinfo
  {journal} {npj Quantum Information}\ }\textbf {\bibinfo {volume} {3}},\
  \bibinfo {eid} {36} (\bibinfo {year} {2017})},\ \Eprint
  {http://arxiv.org/abs/1612.01045} {arXiv:1612.01045 [quant-ph]} \BibitemShut
  {NoStop}%
\bibitem [{\citenamefont {Holland}(1975)}]{Holland75}%
  \BibitemOpen
  \bibfield  {author} {\bibinfo {author} {\bibfnamefont {J.~H.}\ \bibnamefont
  {Holland}},\ }\href {mathworks.com/discovery/genetic-algorithm.html}
  {\enquote {\bibinfo {title} {Adaptation in natural and artificial systems: An
  introductory analysis with applications to biology, control, and artificial
  intelligence.}}\ } (\bibinfo {year} {1975})\BibitemShut {NoStop}%
\end{thebibliography}%


\end{document}